\documentclass[12pt]{article}
\usepackage{amsmath,bbm,amssymb}
\usepackage{amsthm}
\usepackage{comment}
\usepackage{graphicx, graphics}
\usepackage{psfrag,pst-node,subfigure,rotating,amsmath, bbm, amsthm, amssymb, amsthm, setspace, picture, epsfig, amsfonts}

\usepackage{multirow}

\usepackage{arydshln}
\usepackage{algorithm}
\usepackage{algorithmic}

\newtheorem{proposition}{Proposition}

\def\v{\mathbf}\def\vg{\boldsymbol}\def\m{\mathbf}\def\s{\mathcal}

\def\C{{\mathrm{Cov}}}

\begin{document}
\underline{}\begin{center}
    {\Large\bf Personalized Optimization for Computer Experiments}\\{\Large\bf with Environmental Inputs}
\\[4mm] {\large Shifeng \textsc{Xiong}}
\\[2mm] Academy of Mathematics and Systems Science, \\Chinese Academy of Sciences, Beijing 100190\\ Email: xiong@amss.ac.cn
\end{center}

\vspace{1cm} \noindent{\bf Abstract}\quad Optimization problems with both control variables and environmental variables arise in many fields. This paper introduces a framework of personalized optimization to handle such problems.
Unlike traditional robust optimization, personalized optimization devotes to finding a series of optimal control variables for different values of environmental variables. Therefore, the solution from personalized optimization consists of optimal surfaces defined on the domain of the environmental variables. When the environmental variables can be observed or measured, personalized optimization yields more reasonable and better solutions than robust optimization. The implementation of personalized optimization for complex computer models is discussed. Based on statistical modeling of computer experiments, we provide two algorithms to sequentially design input values for approximating the optimal surfaces. Numerical examples show the effectiveness of our algorithms.

\vspace{1cm} \noindent{\bf KEY WORDS:} Expected improvement; Kriging; Personalized medicine; Robust optimization; Sequential design.

\newpage
\section{Introduction}\label{sec:intro}
\hskip\parindent\vspace{-0.8cm}

Optimization has broad applications in engineering, economics, operations research, and many other fields. A canonical form of optimization problems is \begin{equation}\label{opt}\min_{\v{x}\in\s{D}}f(\v{x}),\end{equation}where $f$ is a real-valued function defined on a subset $\s{D}$ of an Euclidean space. Since we use the minimization problem to represent general optimization problems in \eqref{opt}, the objective function $f$ is called \emph{cost} throughout this paper.

In many practical situations, the vector of variables $\v{x}$ in \eqref{opt} consist of two parts: control variables and environmental variables, denoted by $\v{s}$ and $\v{t}$, respectively. Control variables can be set as specific values by an engineer or scientist before the process that yields the value of $f$, while environmental variables, sometimes called noise variables, influence the cost in an uncertain way, depending on the user or on the environment at that time. There are a great number of examples of optimization with both control and environmental variables in quality engineering (Beyer and Sendhoff 2007) as well as in other fields (Ben-Tal, El Ghaoui, and Nemirovski 2009). A common case in medicine is as follows. Suppose that we will evaluate different medical treatments for a certain disease. These treatments can be represented by different values of control variables $\v{s}$. For an individual patient, an index of the treatment effect corresponding to treatment $\v{s}$ is denoted by $f$. Suppose that $f$ also depends on some measurable information, denoted by the covaiates $\v{t}$, of the patient. Here these covariates are actually environmental variables. It is desirable to find the best treatment, and this needs to optimize $f$ with both $\v{s}$ and $\v{t}$.

Since environmental variables $\v{t}$ cannot be controlled, the optimization problem in \eqref{opt} should be modified to accommodate the situations where environmental variables exist. To this end, robust optimization techniques, which find the optimal \emph{robust} value of $\v{s}$ against the uncertainty from $\v{t}$, have been well developed  (Ben-Tal, El Ghaoui, and Nemirovski 2009). There are a number of definitions of this \emph{robustness} with different focuses. In the above medical example, the optimal robust solution can be defined as the treatment that makes the average effects over all patients under investigation the best.

A recent trend in medical research is \emph{personalized medicine} (Jain 2015), also called precision medicine, stratified medicine, or P4 medicine. Roughly speaking, it means selection of treatment best suited for
an individual according to the personal information of the individual, particularly the gene information. Personalized medicine has bred many new statistical issues like subgroup identification which a number of statisticians are working on (Ruberg, Chen, and Wang 2010; Xu et al. 2015; Zhao et al. 2015).
Here we focus on optimization and let us return to the above example where the index $f$ of the treatment effect depends on the covaiates $\v{t}$. Under the spirit of personalized medicine, the best treatment should vary with the change of $\v{t}$, and should optimize $f$ for each given $\v{t}$. It can be expected that the treatment from personalized medicine is generally better than that from robust optimization, which is the same for all $\v{t}$. This idea can be borrowed to cope with general optimization problems with both control and environmental variables. This paper introduces a framework of \emph{personalized optimization}, which devotes to finding a series of optimal control variables for different values of environmental variables. The solution from personalized optimization has a form of several curves or surfaces, called \emph{profile optimal curves} (POC's) or \emph{profile optimal surfaces} (POS's). Personalized optimization is applicable to the situations where the environmental variables can be observed or measured, and seems more reasonable than traditional robust optimization in practice.

When the cost function has an analytic expression, the problem of finding POS's (or POC's) can be solved for given environmental variables by traditional optimization techniques such as the Newton-Raphson algorithm (Lange 2010). In this paper we discuss this problem for computer experiments (Santner, Williams and Notz 2003). Nowadays computer experiments are commonly used to study a computer simulation in engineering and scientific investigations. Computer simulations usually have complex input-output relationships with long running times of the computer codes involved. A computer simulation can be viewed as an expensive black-box function when its output/response is scalar, A popular way to study such expensive function is to build a statistical surrogate model based on a small number of output values on elaborately designed input values (Santner, Williams and Notz 2003). To study optimization algorithms for computer experiments, we need to integrate statistical methods with optimization techniques, and this becomes a hot area in both statistics and optimization because of the broad applications of computer experiments.

When only control inputs exist, Jones, Schonlau, and Welch (1998) proposed an expected improvement (EI) algorithm for optimizing expensive black-box functions. The EI method has proven to be a success, and its variants or related methods are proposed for various cases by many aothors; see Brochu, Cora, and De Freitas (2010), Picheny et al. (2013), and He, Tuo, and Wu (2016) among others. The EI method has also be extended to the situations where environmental inputs exist (Williams, Santner, and Notz 2000; Lehman, Santner, and Notz 2004; Marzat, Walter, and Piet-Lahanier 2013; Marzat, Walter, and Piet-Lahanier 2016). These extended EI methods are all within the framework of robust optimization, and search solutions that optimize certain robust criteria. Unlike them, in this paper we discuss the personalized optimization problem of computer experiments.

This paper is organized as follows. Section \ref{sec:po} introduces some notation and definitions in personalized optimization. Section \ref{sec:med} provides two sequential algorithms to implement personalized optimization for computer experiments. Section \ref{sec:simu} presents numerical examples to evaluate the proposed algorithms. We end the paper with some discussion in Section \ref{sec:dis}.

\section{Personalized optimization: A complement to robust optimization} \label{sec:po}
\hskip\parindent\vspace{-0.8cm}

Consider a cost function \begin{equation}\label{cf}y=f(\v{s},\v{t}),\end{equation}where $\v{s}\in\s{D}_\mathrm{s}\subset{\mathbb{R}}^p$ and $\v{t}\in\s{D}_\mathrm{t}\subset{\mathbb{R}}^q$ are control and environmental variables, respectively. There are two typical robust criteria in robust optimization. The first one is the expectation criterion (Beyer and Sendhoff 2007) \begin{equation*}f_{\mathrm{E}}(\v{s})=\int_{\s{D}_\mathrm{t}}f(\v{s},\v{t})\varphi(\v{t})\,d\,\v{t},\end{equation*}where $\varphi(\v{t})$ is a probability density function on $\s{D}_\mathrm{t}$. The robust solution according to this criterion is \begin{equation}\label{sroi}\v{s}_{\mathrm{E}}^*=\arg\min_{\v{s}\in\s{D}_\mathrm{s}}f_{\mathrm{E}}(\v{s}).\end{equation}A number of authors designed algorithms for approximating $\v{s}_{\mathrm{E}}^*$ in computer experiments (Williams, Santner, and Notz 2000; Janusevskis and Le Riche 2013). The second robust criterion is the maximization criterion (also called worst-case criterion) (Ben-Tal and Nemirovski 1998)\begin{equation*}f_{\mathrm{M}}(\v{s})=\max_{\v{t}\in\s{D}_\mathrm{t}}f(\v{s},\v{t}).\end{equation*} Denote the corresponding solution by
\begin{equation}\label{srom}\v{s}_{\mathrm{M}}^*=\arg\min_{\v{s}\in\s{D}_\mathrm{s}}f_{\mathrm{M}}(\v{s}).\end{equation}Research articles on this criterion in computer experiments include Marzat, Walter, and Piet-Lahanier (2013) and Marzat, Walter, and Piet-Lahanier (2016), etc. There are other robust criteria used in practice; see Beyer and Sendhoff (2007).

When $\v{t}$ can be observed or measured, the optimal $\v{s}$ should minimize the cost for each given $\v{t}$, and this is the main idea of personalized optimization. In other words, the purpose of personalized optimization is to find the \emph{profile optimal surfaces} (POS's) (or called \emph{profile optimal curves} (POC's) for $q=1$)\begin{equation}\label{pos}\mathfrak{s}(\v{t})=\arg\min_{\v{s}\in\s{D}_\mathrm{s}}f(\v{s},\v{t}).\end{equation}
The right side of \eqref{pos} may be non-unique. For such cases, we assign any minimum to define $\mathfrak{s}(\v{t})$. We call a mapping from $\s{D}_\mathrm{t}$ to $\s{D}_\mathrm{s}$ \emph{personalized decision} in the following. Clearly, $\mathfrak{s}$ in \eqref{pos} is a personalized decision. In addition, the robust solutions $\v{s}_{\mathrm{E}}^*$ in \eqref{sroi} and $\v{s}_{\mathrm{M}}^*$ in \eqref{srom} actually define two constant personalized decisions
\begin{equation}\label{uem}u_{\mathrm{E}}(\v{t})\equiv \v{s}_{\mathrm{E}}^*\quad\text{and}\quad u_{\mathrm{M}}(\v{t})\equiv \v{s}_{\mathrm{M}}^*,\end{equation}respectively.
Since for any $\v{s}$ and $\v{t}$, \begin{equation}\label{ine}f(\mathfrak{s}(\v{t}),\v{t})\leqslant f(\v{s},\v{t}),\end{equation} $\mathfrak{s}$ is the best personalized decision in the sense of minimizing the cost.

For a personalized decision $u$, define its expected cost and maximum cost as \begin{eqnarray}&&C_{\mathrm{E}}(u)=\int_{\s{D}_\mathrm{t}}f(u(\v{t}),\v{t})\varphi(\v{t})\,d\,\v{t},\label{ce}\\&&C_{\mathrm{M}}(u)=\max_{\v{t}\in\s{D}_\mathrm{t}}f(u(\v{t}),\v{t}),\label{cm}\end{eqnarray}respectively. The expected cost and maximum cost can be used to evaluate the overall performance of $u$. We then use them to compare robust optimization and personalized optimization.
By \eqref{ine} and the definitions in \eqref{uem}, we have\begin{equation*}C_{\mathrm{E}}(\mathfrak{s})\leqslant C_{\mathrm{E}}(u_{\mathrm{E}})\leqslant C_{\mathrm{E}}(u_{\mathrm{M}})\end{equation*}and \begin{equation*}C_{\mathrm{M}}(\mathfrak{s})\leqslant C_{\mathrm{M}}(u_{\mathrm{M}})\leqslant C_{\mathrm{M}}(u_{\mathrm{E}}).\end{equation*}
Consider a simple example with $f(s,t)=(s-t)^2,\ s,t\in[0,1]$. Some algebra gives that $C_{\mathrm{E}}(\mathfrak{s})=C_{\mathrm{M}}(\mathfrak{s})=0$, $C_{\mathrm{E}}(u_{\mathrm{E}})=1/12$, and $C_{\mathrm{M}}(u_{\mathrm{M}})=1/4$.

\section{Sequential algorithms for searching POS's of computer models} \label{sec:med}
\hskip\parindent\vspace{-8mm}

This section discusses personalized optimization for computer experiments. Since computer models are often expensive black-box functions, we first describe the popular statistical modeling methods for them in Section \ref{subsec:krig}. In Section \ref{subsec:alg} we present two sequential algorithms to search POS's of computer models.

\subsection{The Gaussian process model} \label{subsec:krig}
\hskip\parindent\vspace{-0.8cm}

The Gaussian process model (also called Kriging model) (Matheron 1963) is widely used to analyze computer experiments (Sacks et al. 1989). It models the output of a computer simulation as 
\begin{equation}f(\v{x})=\v{g}(\v{x})'\vg{\beta}+Z(\v{x}),\label{kriging}\end{equation}where $\v{x}\in\s{D}\subset \mathbb{R}^d$, $\v{g}(\v{x})=\left(g_1(\v{x}),\ldots,g_m(\v{x})\right)'$ is
a pre-specified set of functions, $\vg{\beta}$ is a vector of unknown regression coefficients, and $Z(\v{x})$ is a stationary
Gaussian process with mean zero, variance $\sigma^2$, and correlation parameters $\vg{\theta}$, denoted by GP$(0, \sigma^2, \vg{\theta})$. The covariance between
$Z(\v{x}_1)$ and $Z(\v{x}_2)$ in \eqref{kriging} is represented by \begin{equation}\C[Z(\v{x}_1),
Z(\v{x}_2)]=\sigma^2R(\v{x}_1-\v{x}_2\,|\,\vg{\theta}),\label{R}\end{equation} where $R(\cdot\,|\,\vg{\theta})$ is the correlation function depending on a parameter
vector $\vg{\theta}$. A popular choice of $R$ is the Gaussian correlation function
\begin{eqnarray}
R(\v{u}\,|\,\theta)=\exp(-\theta_i u_i^2)\quad\text{for}\ \v{u}=(u_1,\ldots,u_d)'\in\s{D},\label{GR}
\end{eqnarray} where $\theta_1,\ldots,\theta_d>0$ are the correlation parameters.

The parameters in model \eqref{kriging} can be estimated by the maximum likelihood method. Suppose the set of input values is $\{\v{x}_1,\ldots,\v{x}_n\}$, where
$\v{x}_i=(x_{i1},\ldots,x_{id})'$ for $i=1,\ldots,n$. The corresponding response values $\v{y}=(f(\v{x}_1),\ldots,f(\v{x}_n))'$. The negative log-likelihood, up to an additive constant, is proportional to
\begin{eqnarray}n\log(\sigma^2)+\log({\mathrm{det}}(\m{R}))+(\v{y}-\m{G}\vg{\beta})'\m{R}^{-1}(\v{y}-\m{G}\vg{\beta})/\sigma^2,\label{ll}\end{eqnarray}
where $\m{R}$ is the $n\times n$ correlation matrix whose $(i,j)$th entry is $R(\v{x}_i-\v{x}_j\,|\,\vg{\theta})$ defined in (\ref{R}), ``${\mathrm{det}}$" denotes
matrix determinant, and $\m{G}=\left(\v{g}(\v{x}_1),\ldots,\v{g}(\v{x}_n)\right)'$.

When $\vg{\theta}$ is known, the maximum likelihood estimators (MLEs) of $\vg{\beta}$ and $\sigma^2$ are
\begin{equation}\left\{\begin{array}{l}\hat{\vg{\beta}}=(\m{G}'\m{R}^{-1}\m{G})^{-1}\m{G}'\m{R}^{-1}\v{y},
\\\hat{\sigma}^2=(\v{y}-\m{G}\hat{\vg{\beta}})'\m{R}^{-1}(\v{y}-\m{G}\hat{\vg{\beta}})/n.\end{array}\right. \label{bs}\end{equation}
For an untried point $\v{x}_0\in\s{D}$, the best linear unbiased predictor $\hat{f}$ of $f$ (Santner, Williams and Notz 2003) is
\begin{eqnarray}\hat{f}(\v{x}_0)=\v{g}(\v{x}_0)'\hat{\vg{\beta}}+{\v{r}_0}'{\m{R}}^{-1}\big(\v{y}-\m{G}\hat{\vg{\beta}}\big),\label{blup}\end{eqnarray}
where ${\v{r}_0}=\big(R(\v{x}_0-\v{x}_1\,|\,{\vg{\theta}}),\ldots,R(\v{x}_0-\v{x}_n\,|\,{\vg{\theta}})\big)'$.
Given $\alpha\in(0,1)$, the $(1-\alpha)$ prediction interval of $f(\v{x}_0)$ is\begin{equation}\label{pi}P\left(f(\v{x}_0)\in \hat{f}(\v{x}_0)\pm \phi(\v{x}_0)\,t_{n-d}(\alpha/2)\right)=1-\alpha,\end{equation} where
$\phi(\v{x}_0)\geqslant0$,\begin{eqnarray}&&\phi(\v{x}_0)^2=\frac{Q^2}{n-d}\left\{1-(\v{g}(\v{x}_0)',\v{r}_0')\left(\begin{array}{cc}\m{0}&\m{G}'\\\m{G}&\m{R}\end{array}\right)^{-1}
\left(\begin{array}{c}\v{g}(\v{x}_0)\\\v{r}_0\end{array}\right)\right\},\label{phi}\\&&Q^2=\v{y}'\big[\m{R}^{-1}-\m{R}^{-1}\m{G}(\m{G}'\m{R}^{-1}\m{G})^{-1}\m{G}'\m{R}^{-1}\big]\v{y},\nonumber\end{eqnarray}and $t_{n-d}(\alpha/2)$ is the upper $\alpha/2$ quantile of Student's $t$-distribution with $n-d$ degrees of freedom. Denote the lower bound in \eqref{pi} by \begin{equation}\label{lb}L(\v{x}_0)=\hat{f}(\v{x}_0)-{\phi(\v{x}_0)}\,t_{n-d}(\alpha/2).\end{equation}
When $\vg{\theta}$ is unknown, by plugging \eqref{bs} into \eqref{ll}, we have the MLE of $\vg{\theta}$ 
\begin{equation*}\hat{\vg{\theta}}=\arg\min_{\vg{\theta}}\,n\log(\hat{\sigma}^2)+\log\left({\mathrm{det}}({\m{R}})\right).
\end{equation*}The predictor $\hat{f}$ in \eqref{blup} and the prediction intervals in \eqref{pi} can be modified by replacing $\vg{\theta}$ with $\hat{\vg{\theta}}$.

Let us return to the computer model having the form in \eqref{cf}, where $\v{x}$ in \eqref{kriging} is $(\v{s}',\v{t}')'$, $d=p+q$, and $\s{D}=\s{D}_\mathrm{s}\times\s{D}_\mathrm{t}$. Suppose that we have $n$ pairs of input values $(\v{s}_1',\v{t}_1')',\ldots,(\v{s}_n',\v{t}_n')'$, and the corresponding output values $\v{y}=\left(f(\v{s}_1,\v{t}_1),\ldots,f(\v{s}_n,\v{t}_n)\right)'$. By \eqref{blup}, the predictor $\hat{f}$ can be obtained based on the data. Consequently, the POS's can be estimated by \begin{equation}\label{nai}\hat{\mathfrak{s}}(\v{t})=\arg\min_{\v{s}\in\s{D}_\mathrm{s}}\hat{f}(\v{s},\v{t}).\end{equation}

\subsection{Algorithm description} \label{subsec:alg}
\hskip\parindent\vspace{-0.8cm}

This subsection presents two sequential algorithms to search POS's. Since more experimental points closed to the true POS's can make their estimators in \eqref{nai} more accurate, the basic idea of our algorithms is to sequentially select the next point according to a criterion that measures the proximity of a point to the POS's.

First we give two useful definitions. For $\v{t}\in\s{D}_\mathrm{t}$, define \begin{equation}\label{th}\tilde{\mathfrak{s}}(\v{t})=\arg\min_{\v{s}\in\s{D}_\mathrm{s}}L(\v{s},\v{t}),\end{equation}where $L$ is defined in \eqref{lb}. For a set $\s{X}$ of points, define\begin{equation}\label{mf}M(\v{x},\s{X})=\min_{\v{a}\in\s{X}}\|\v{x}-\v{a}\|.\end{equation}The proposed algorithms are hierarchical. Suppose that the current experimental points are $(\v{s}_1',\v{t}_1')',\ldots,(\v{s}_{n}',\v{t}_{n}')'$. For given $\v{t}_{n+1}$, the goal is to optimize $f$ with respect to $\v{s}$. Hence the idea of EI (Jones, Schonlau, and Welch 1998) can be used. Unfortunately, since there is no, or only a part of, experimental points on the profile $\{(\v{s}',\v{t}')'\in\s{D}:\ \v{t}=\v{t}_{n+1}\}$, the EI criterion is not well defined for our problem. Here we adopt the confidence bound criterion (Cox and John 1997) in \eqref{th} to find the next $\v{s}$, that is, to use $\v{s}_{n+1}$ that minimizes the lower bound of the $(1-\alpha)$ prediction interval of $f(\v{s},\v{t}_{n+1})$. The confidence bound criterion takes the uncertainty in predicting the response on untried points into account like EI, and has a simple form for our problem. The nominal level $\alpha$ controls the selection between the global solution and local solution. For smaller value of $\alpha$, greater weight in the confidence bound criterion will be placed on the variance of the predictor, and thus we tend to approximate the global minimizer. Such a search, however, can be slow to reach a local solution. Therefore, smaller $\alpha$ is more suitable for multimodal functions.

We next discuss the selection of $\v{t}_{n+1}$. Two selection methods are presented, which form the difference between the two proposed algorithms SHA1 and SHA2; see Algorithms \ref{ag:sha1} and \ref{ag:sha2}. The first method is model free through optimizing a space-filling criterion in \eqref{mf}. This criterion stems from the maximin distance criterion (Johnson, Moore, and Ylvisaker 1990). The second method is model-based; see Step 3 in SHA2. It selects the next $\v{t}$ with the maximal amount of uncertainty at the point which consists of $\v{t}$ and $\v{s}$ selected from the confidence bound criterion. We will compare SHA1 and SHA2 with different $\alpha$ via numerical examples in the next section.

\renewcommand{\algorithmicrequire}{\textbf{Inputs:}}
\renewcommand{\algorithmicensure}{\textbf{Steps:}}
\floatname{algorithm}{Algorithm}
\begin{algorithm}[htb]
\caption{\label{ag:sha1}\quad The sequential hierarchical algorithm 1 (SHA1)}
\begin{algorithmic}[1]
\REQUIRE ~~\\  The initial sample size $n_0$ and the coverage level $\alpha\in(0,1)$. \\[1mm]\ENSURE ~~\\ 1. Choose the initial design $\s{X}=\{(\v{s}_1',\v{t}_1')',\ldots,(\v{s}_{n_0}',\v{t}_{n_0}')'\}\subset\s{D}_\mathrm{s}\times\s{D}_\mathrm{t}$ and compute the corresponding responses $\v{y}=(y_1,\ldots,y_{n_0})'$. Let $n=n_0$.
\\[1mm]\hspace{-4mm} For $n=n_0,n_0+1,\ldots,$\\2. Fit the Gaussian process model with the current data $\{\s{X},\v{y}\}$.
\\3. Find $$\v{t}_{n+1}=\arg\max_{\v{t}\in\s{D}_\mathrm{t}}M\big(\v{t},\s{T}_n\big),$$
where $\s{T}_n=\{\v{t}_1,\ldots,\v{t}_n\}$ and $M$ is defined in \eqref{mf}. Let $\v{s}_{n+1}=\tilde{\mathfrak{s}}(\v{t}_{n+1})$.
\\4. Compute $y_{n+1}=f(\v{s}_{n+1},\v{t}_{n+1})$.
\\5. Let $\s{X}\leftarrow\s{X}\cup\{(\v{s}_{n+1}',\v{t}_{n+1}')'\}$, $\v{y}\leftarrow(\v{y}',y_{n+1})'$, and $n\leftarrow n+1$.\\\hspace{-4mm}  End for if some stopping rule is met.
\end{algorithmic}
\end{algorithm}

\renewcommand{\algorithmicrequire}{\textbf{Inputs:}}
\renewcommand{\algorithmicensure}{\textbf{Steps:}}
\floatname{algorithm}{Algorithm}
\begin{algorithm}[htb]
\caption{\label{ag:sha2}\quad The sequential hierarchical algorithm 2 (SHA2)}
\begin{algorithmic}[1]
\REQUIRE ~~The same as in Algorithm \ref{ag:sha1}. \\[1mm]\ENSURE ~~\\
All steps are the same as in Algorithm \ref{ag:sha1} except Step 3: Find $$\v{t}_{n+1}=\arg\max_{\v{t}\in\s{D}_\mathrm{t}}\phi\big(\tilde{\mathfrak{s}}(\v{t}),\v{t}\big),$$
where $\phi$ and $\tilde{\mathfrak{s}}$ are defined in \eqref{phi} and \eqref{th}, respectively.
Let $\v{s}_{n+1}=\tilde{\mathfrak{s}}(\v{t}_{n+1})$.
\end{algorithmic}
\end{algorithm}

In SHA1 and SHA2, the initial design can be chosen as a space-filling design (Joseph 2016) which are commonly used in computer experiments. Such designs include the maximin distance design (Johnson,
Moore, and Ylvisaker 1990), Latin hypercube design (Mckay, Beckman, and Conover 1979), and uniform design (Fang et al. 2000). We use the Gaussian correlation function \eqref{GR} to build the Gaussian process model in \eqref{kriging} throughout, where $\v{g}(\v{x})$ in \eqref{kriging} is set as $(1,s_1,\ldots,s_p,t_1,\ldots,t_q)'$.
Let $\hat{f}_n$ denote the predictor by \eqref{blup} based on the current data in the $n$th iteration, and then the estimators of POS's, $\hat{\mathfrak{s}}_n(\v{t})$, can be computed by \eqref{nai}. We stop the iterations in SHA1 and SHA2 if
$$\int_{\s{D}_\mathrm{t}}\frac{|\hat{f}_{n-1}(\hat{\mathfrak{s}}_{n-1}(\v{t}),\v{t})-\hat{f}_{n}(\hat{\mathfrak{s}}_{n}(\v{t}),\v{t})|}{|\hat{f}_{n}(\hat{\mathfrak{s}}_{n}(\v{t}),\v{t})|}\varphi(\v{t})\,d\,\v{t}
<\varepsilon_1\quad\text{and}\quad \int_{\s{D}_\mathrm{t}}\phi(\hat{\mathfrak{s}}_{n}(\v{t}),\v{t})\varphi(\v{t})\,d\,\v{t}<\varepsilon_2,$$where $\varepsilon_1$ and $\varepsilon_2$ are pre-specified thresholdings.
The stopping rule can also be set as$$\max_{\v{t}\in\s{D}_\mathrm{t}}\frac{|\hat{f}_{n-1}(\hat{\mathfrak{s}}_{n-1}(\v{t}),\v{t})-\hat{f}_{n}(\hat{\mathfrak{s}}_{n}(\v{t}),\v{t})|}{|\hat{f}_{n}(\hat{\mathfrak{s}}_{n}(\v{t}),\v{t})|}
<\varepsilon_1\quad\text{and}\quad \max_{\v{t}\in\s{D}_\mathrm{t}}\phi(\hat{\mathfrak{s}}_{n}(\v{t}),\v{t})<\varepsilon_2.$$
Besides, a practical and simple stopping rule in SHA1 and SHA2 is to stop the iterations until the run size reaches our budget.

\section{Numerical illustrations} \label{sec:simu}
\hskip\parindent\vspace{-1.6cm}

\subsection{A simple two-dimensional example} \label{subsec:2d}
\hskip\parindent\vspace{-0.8cm}

\begin{figure}[t]\begin{center}\scalebox{0.6}[0.6]{\includegraphics{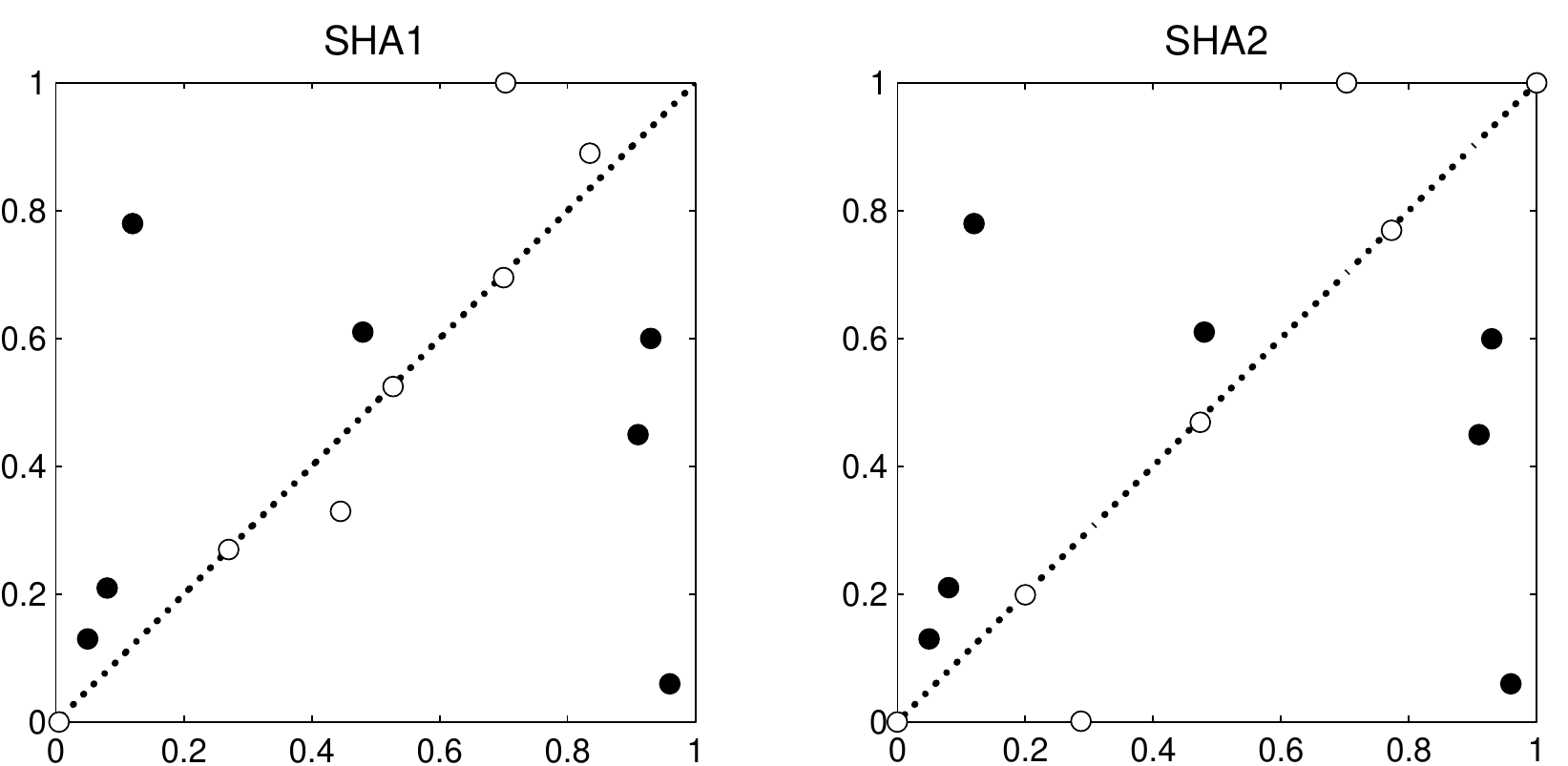}}\end{center}
\caption{The designs corresponding to SHA1 and SHA2 ($\alpha=0.2$) in Section \ref{subsec:2d}. The black dots constitute the initial design, and the white dots denote the added runs after seven iterations. The dotted line is the true POC.}\label{fig:2d}
\end{figure}

Consider the simple function $f(s,t)=(s-t)^2,\ s,t\in[0,1]$, which is used in Section \ref{sec:po}. It is obvious that the POC of $f$ is $s=t,\ t\in[0,1]$. We begin with seven randomly generated points, and use SHA1 and SHA2 to approximate the POC. Figure \ref{fig:2d} shows the locations of the added runs after seven iterations. We can see that these points are closed to the POC, and this indicates that our algorithms work well.

\subsection{Comparison of the costs from different methods} \label{subsec:com}
\hskip\parindent\vspace{-0.8cm}

Let $\s{D}_\mathrm{s}=[0,1]^p$, $\s{D}_\mathrm{t}=[0,1]^q$, and $\varphi$ in \eqref{ce} be the density of the uniform distribution on $[0,1]^q$. We use the following six test functions:
\begin{eqnarray*}
&&f_1(s,t)=2|s^3-t|+\exp(t)(s-2t)^2,
\\&&f_2(s,t)=\cos(10\sqrt{s^2+t^2})/(\sqrt{s^2+t^2}+1),
\\&&f_3(s,t)=\min\{3-2s+3t,\ 3+2s-t\},
\\&&f_4(s,t)=\left\{15t-\frac{5.1}{4\pi^2}(15s-5)^2+\frac{5}{\pi}(15s-5)-6\right\}^2+10\left(1-\frac{1}{8\pi}\right)\cos(15s-5)+10,
\\&&f_5(\v{s},\v{t})=\left(s_1-|t_1-t_2|\right)^2+\left(s_2-\sqrt{(t_1^2+t_2^2)/2}\right)^4,\quad (p=q=2)
\\&&f_6(\v{s},\v{t})=\sin(5s_1^2)(t_1+2s_2)-\cos(5s_3^2)/\sqrt{1+s_4^2}-2t_2(s_1-s_4),\quad (p=4,q=2)
\end{eqnarray*}some of which have been used in the literature; see Jones, Schonlau, and Welch (1998) and Marzat, Walter, and Piet-Lahanier (2016), among others.
We compare three sequential design strategies, SHA1, SHA2, and the Sobol' sequence (Niederreiter 1992) in approximation to POS's. The Sobol' sequence can be viewed as a sequential uniform design, which places the points more and more uniformly over the experimental region as the number of points increases. In SHA1 and SHA2, the initial design is generated by the first $n_0$ points of the Sobol' sequence, where $n_0=10$ for $f_1,\ldots,f_4$ and $n_0=20$ for $f_5$ and $f_6$. Three nominal levels, $\alpha=0.2,\ 0.5,$ and $0.8$, are used in SHA1 and SHA2. The expected costs in \eqref{ce} and the maximum costs in \eqref{cm} of the estimated POS's, $\hat{\mathfrak{s}}$ in \eqref{nai}, based on the three design strategies are computed in each iteration for the comparison. To compare personalized optimization and robust optimization, we also compute the expected and maximum costs of the constant personalized decisions
$u_{\mathrm{E}}(\v{t})$ and $u_{\mathrm{M}}(\v{t})$ in \eqref{uem} corresponding to the solutions from robust optimization. Here $u_{\mathrm{E}}(\v{t})$ and $u_{\mathrm{M}}(\v{t})$ are accurately computed by the forms of the cost functions. Therefore, they are better than their data-based estimators.

\begin{figure}[p]\begin{center}\scalebox{0.4}[0.6]{\includegraphics{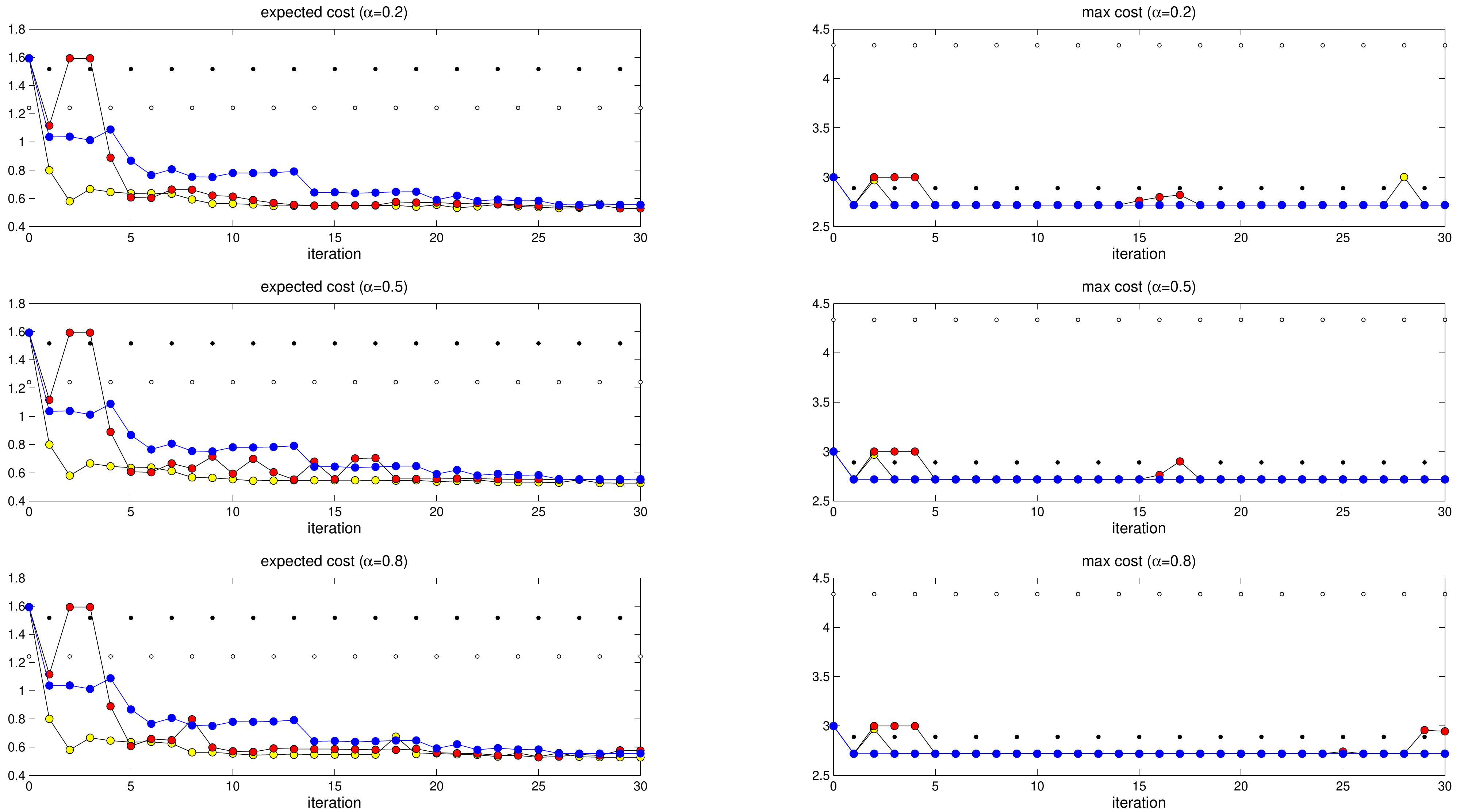}}\end{center}
\caption{The costs concerning $f_1$ in Section \ref{subsec:com}. Legend: white dots --- $u_{\mathrm{E}}$, black dots --- $u_{\mathrm{M}}$, blue dots --- Sobol' sequence, red dots --- SHA1, yellow dots --- SHA2.}\label{fig:fun1}
\end{figure}

\begin{figure}[p]\begin{center}\scalebox{0.4}[0.6]{\includegraphics{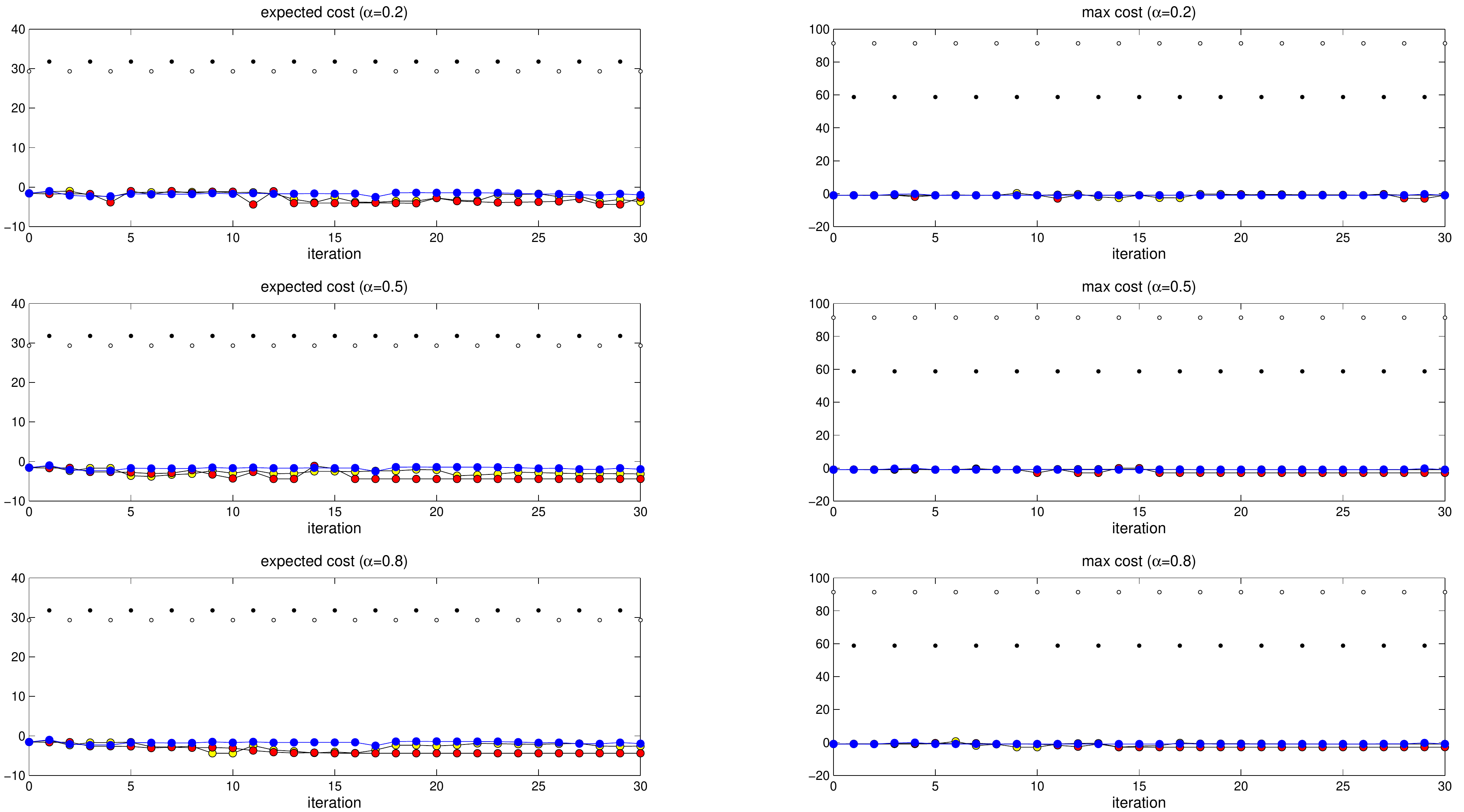}}\end{center}
\caption{The costs concerning $f_6$ in Section \ref{subsec:com}. Legend: white dots --- $u_{\mathrm{E}}$, black dots --- $u_{\mathrm{M}}$, blue dots --- Sobol' sequence, red dots --- SHA1, yellow dots --- SHA2.}\label{fig:fun6}
\end{figure}

The numerical results for 30 iterations are presented in Figures \ref{fig:fun1}\,--\,\ref{fig:fun6}. We can see that, the three methods for approximating POS's are better than the constant personalized decisions in most cases. In addition, SHA1 and SHA2 with larger $\alpha$ are generally better than the Sobol' sequence. For the two proposed algorithms, SHA2 seems to have slightly better overall performance than SHA1. The numerical results indicate that, first, the methods of personalized optimization, even not accurate, are generally better than the accurate solutions of robust optimization; second, the sequential algorithms designed for searching POS's can approximate the POS's better than space-filling designs.

\begin{figure}[p]\begin{center}\scalebox{0.4}[0.6]{\includegraphics{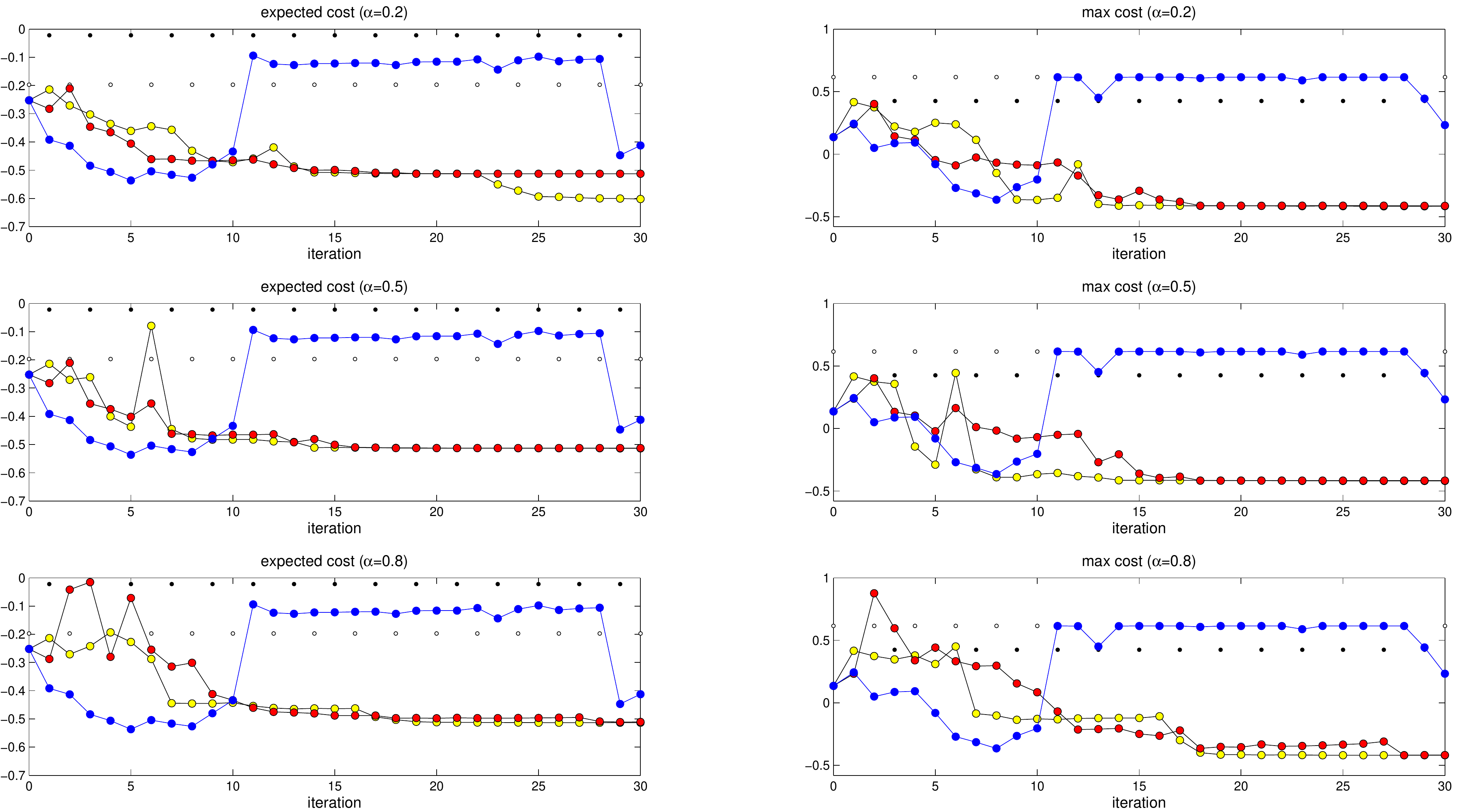}}\end{center}
\caption{The costs concerning $f_2$ in Section \ref{subsec:com}. Legend: white dots --- $u_{\mathrm{E}}$, black dots --- $u_{\mathrm{M}}$, blue dots --- Sobol' sequence, red dots --- SHA1, yellow dots --- SHA2.}\label{fig:fun2}
\end{figure}

\begin{figure}[p]\begin{center}\scalebox{0.4}[0.6]{\includegraphics{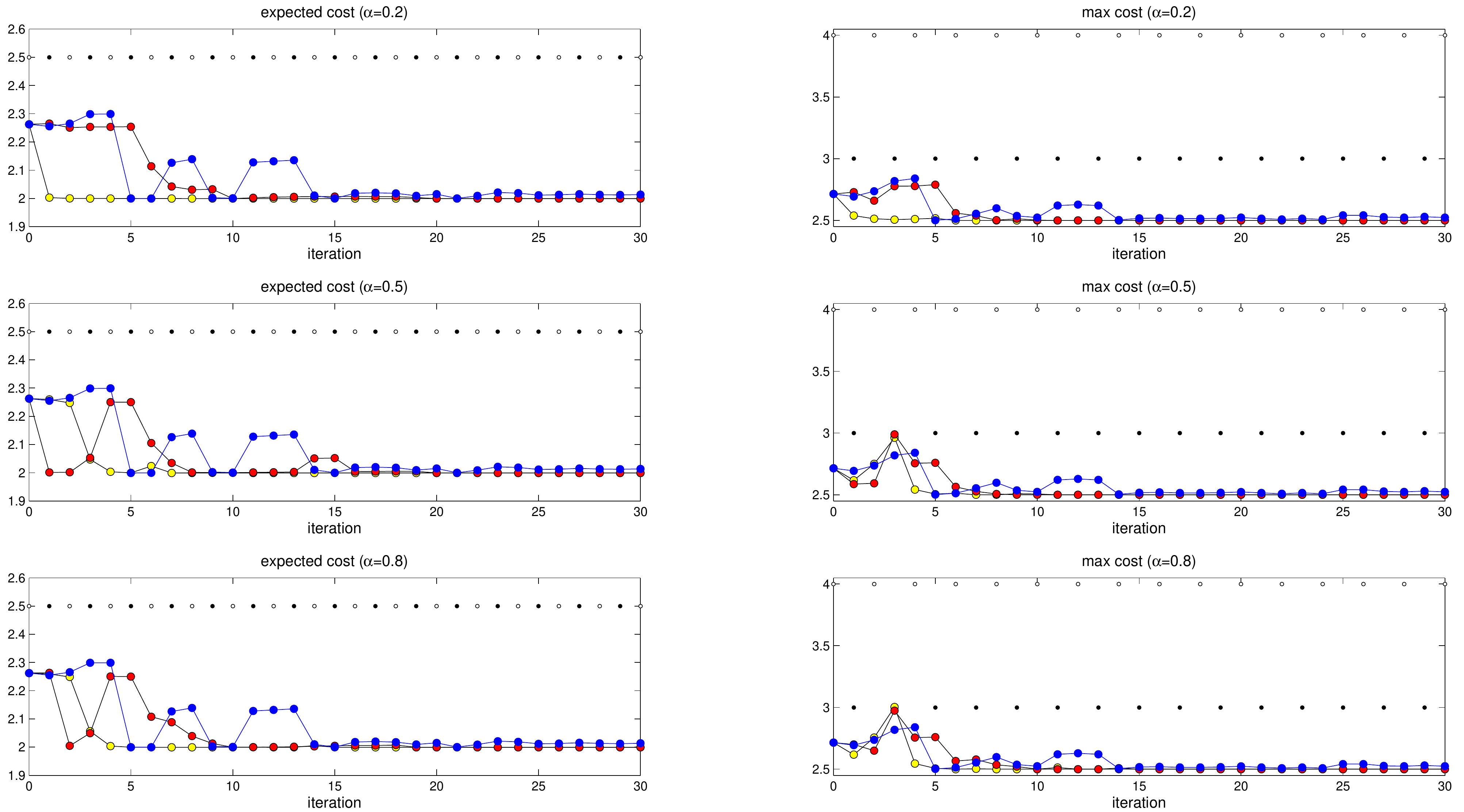}}\end{center}
\caption{The costs concerning $f_3$ in Section \ref{subsec:com}. Legend: white dots --- $u_{\mathrm{E}}$, black dots --- $u_{\mathrm{M}}$, blue dots --- Sobol' sequence, red dots --- SHA1, yellow dots --- SHA2.}\label{fig:fun3}
\end{figure}

\begin{figure}[p]\begin{center}\scalebox{0.4}[0.6]{\includegraphics{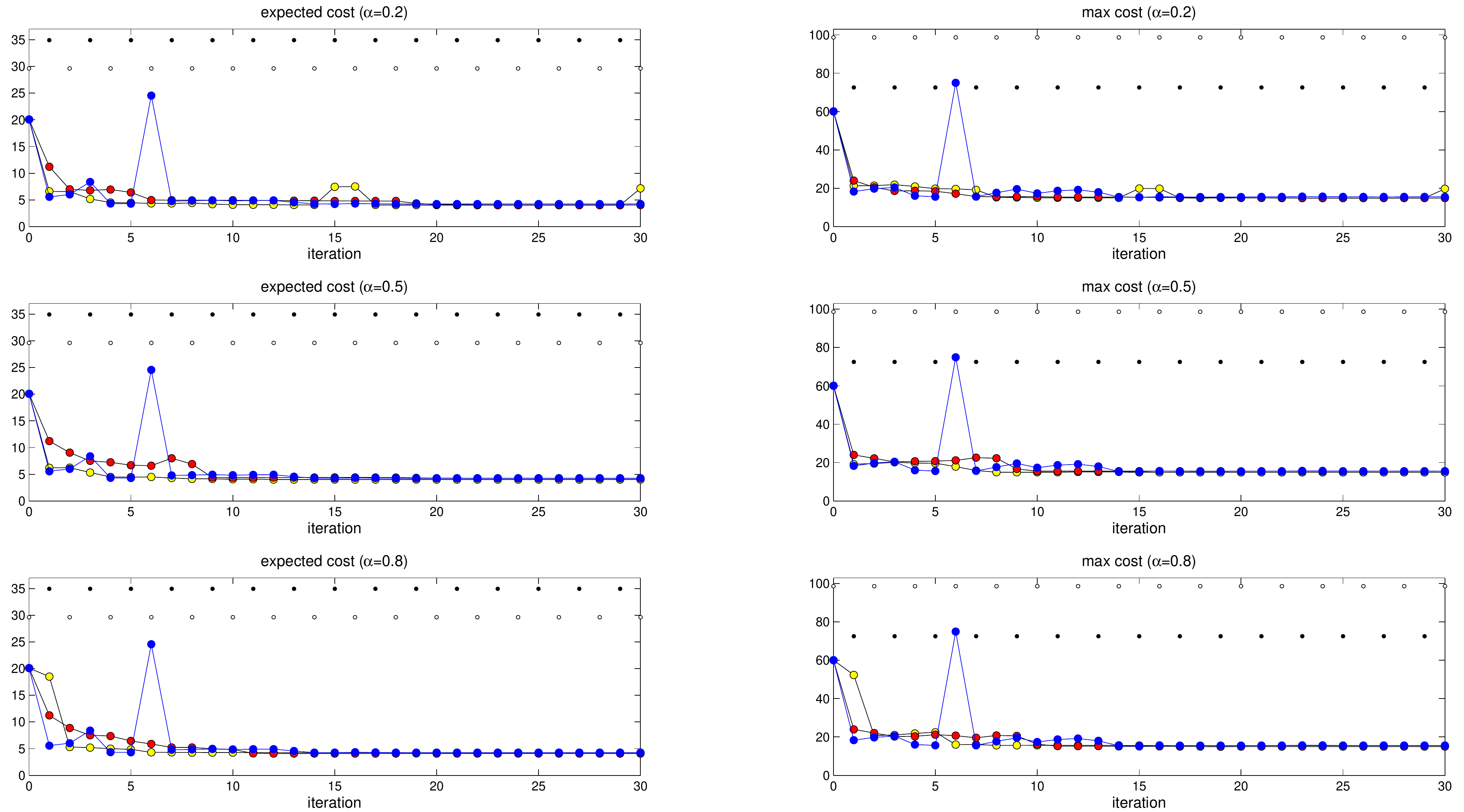}}\end{center}
\caption{The costs concerning $f_4$ in Section \ref{subsec:com}. Legend: white dots --- $u_{\mathrm{E}}$, black dots --- $u_{\mathrm{M}}$, blue dots --- Sobol' sequence, red dots --- SHA1, yellow dots --- SHA2.}\label{fig:fun4}
\end{figure}

\begin{figure}[p]\begin{center}\scalebox{0.4}[0.6]{\includegraphics{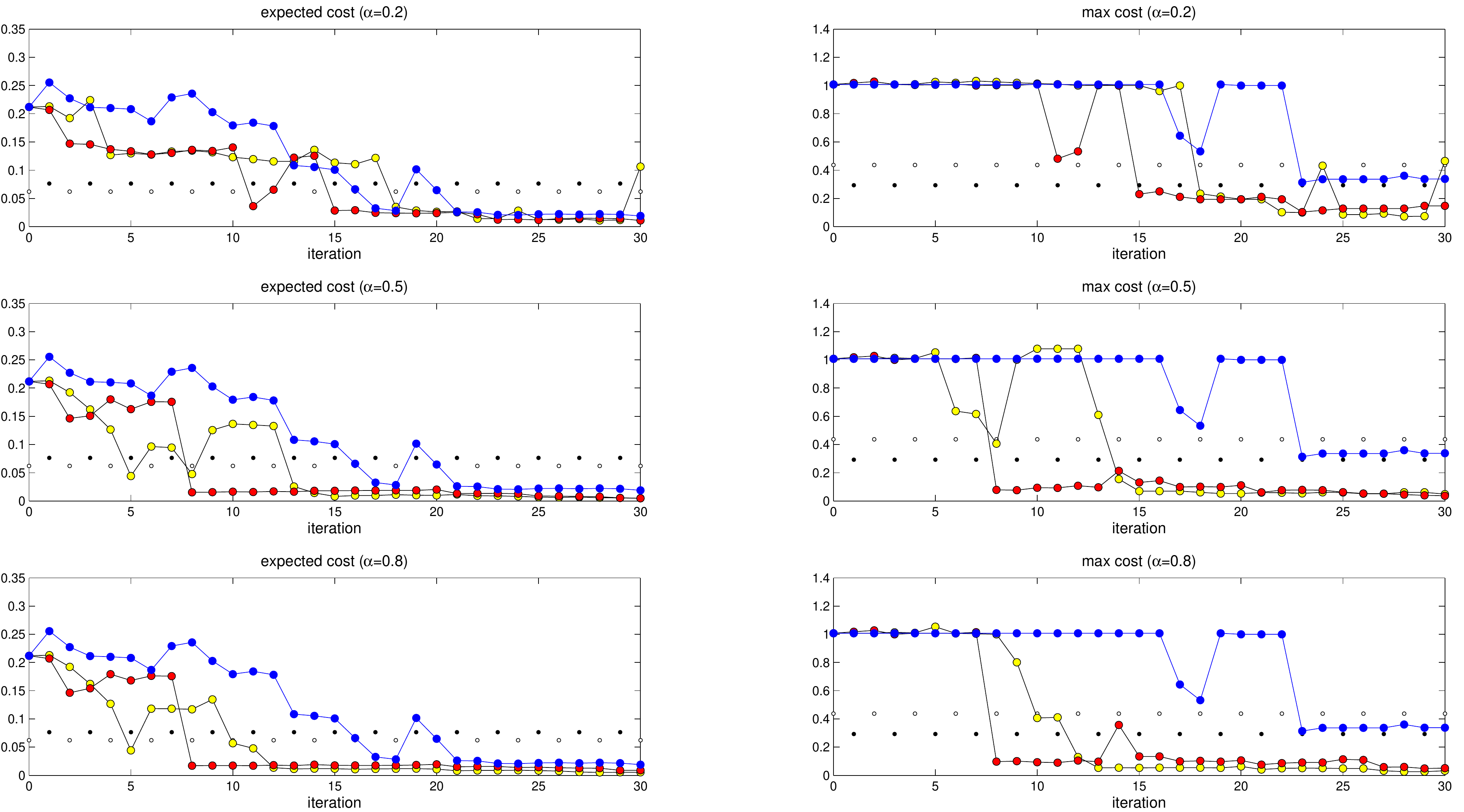}}\end{center}
\caption{The costs concerning $f_5$ in Section \ref{subsec:com}. Legend: white dots --- $u_{\mathrm{E}}$, black dots --- $u_{\mathrm{M}}$, blue dots --- Sobol' sequence, red dots --- SHA1, yellow dots --- SHA2.}\label{fig:fun4}
\end{figure}

\section{Discussion} \label{sec:dis}
\hskip\parindent\vspace{-0.8cm}

This paper has introduced the framework of personalized optimization to handle optimization problems with both control variables and environmental variables. We have discussed its implementation for computer experiments. Two sequential algorithms for searching POS's of complex computer models have been proposed. Theoretical and numerical analysis indicates that, when the environmental variables can be observed or measured, personalized optimization yields more reasonable and better solutions than traditional robust optimization.

Personalized optimization for computer experiments can be studied further in several directions. First, for large $p$ and/or $q$, it is difficult to solve the optimization problems that are needed in personalized optimization such as those in \eqref{nai} and Step 3 in SHA1. This is a common difficulty in optimization. Some global optimization methods can be tried, including multiple-startpoint methods (Ugray et al. 2007) and random search algorithms (Kirkpatrick, Gelatt, and Vecchi 1983; Dorsey and Mayer 1995). Sensitivity analysis techniques (Saltelli et al. 2008) can also be used to reduce the dimensionality of the initial optimization problems. Second, we can use different nominal level $\alpha$ for different $\v{t}$ in SHA1 and SHA2. This modification may be better when the cost function has different levels of multimodality for different $\v{t}$. Third, the proposed algorithms can be extended to the cases where both measurable and unmeasurable environmental variables exist. We need to combine our methods with robust optimization techniques for such cases. Fourth, since SHA1 and SHA2 are proposed to search optimal surfaces, there is not straightforward mathematical tool to study their convergence. It is valuable to develop a theoretical framework to fill this gap in the future.


\section*{Acknowledgements}

This work is supported by the National Natural Science Foundation of China (Grant No. 11271355, 11471172) and Key Laboratory of Systems and Control, Chinese Academy of Sciences.


\vspace{1cm}
\begin{center}{\large REFERENCES}\end{center}

{\footnotesize
\begin{description}

\item{}
Ben-Tal, A., El Ghaoui, L., and Nemirovski, A. (2009). \textit{Robust Optimization}. Princeton University Press: Princeton and Oxford.

\item{}
Ben-Tal, A. and Nemirovski, A. (1998). Robust convex optimization. \textit{Mathematics of Operations Research}, 23, 769--805.

\item{}
Beyer, H. G. and Sendhoff, B. (2007). Robust optimization: a comprehensive survey. \textit{Computer methods in applied mechanics and engineering}, 196, 3190--3218.

\item{}
Brochu, E., Cora, V. M., and De Freitas, N. (2010). A tutorial on Bayesian optimization of expensive cost functions, with application to active user modeling and hierarchical reinforcement learning. arXiv preprint arXiv:1012.2599.




\item{}
Cox, D. D. and John, S., (1997). SDO: A statistical method for global optimization. In M. N. Alexandrov and M. Y. Hussaini, editors, \textit{Multidisciplinary Design Optimization: State of the Art}, pages 315--329. SIAM.



\item{}
Dorsey, R. E. and Mayer, M. J. (1995). Genetic algorithms for estimation problems with multiple optima, nondifferentiability, and other irregular features. \textit{Journal of Business and Economic Statistics}, 13, 53--66.

\item{}
Fang, K. T., Lin, D. K. J., Winker, P.. and Zhang, Y. (2000). Uniform design: theory and application. \textit{Technometrics}, 42, 953--963.


\item{}
He, X., Tuo, R., and Wu, C. F. J. (2016). Optimization of multi-fidelity computer experiments via the EQIE criterion, \textit{Technometrics}, In press.


\item{}
Jain, K. K. (2015). \textit{Textbook of Personalized Medicine}, second edition. New York: Springer.

\item{}
Janusevskis, J. and Le Riche, R. (2013): Simultaneous Kriging-based estimation and optimization of mean response. \textit{Journal of Global optimization}, 55, 313--336.


\item{}
Johnson, M. E., Moore, L. M., and Ylvisaker, D. (1990). Minimax and maximin distance designs. \textit{Journal of Statistical Planning and Inference}, 26, 131--148.

\item{}
Jones, D. R., Schonlau, M., and Welch, W. J. (1998). Efficient global optimization of expensive black-box functions. \textit{Journal of Global optimization}, 13, 455-492.

\item{}
Joseph, V. R. (2016). Space-filling designs for computer experiments: a review, \textit{Quality Engineering}, 28, 28--35.



\item{}
Kirkpatrick, S., Gelatt, C. D., and Vecchi, M. P. (1983). Optimization by simulated annealing. \textit{Science}, 220, 671--680.

\item{}
Lange, K. (2010). \textit{Numerical Analysis for Statisticians}. Springer Science \& Business Media.

\item{}
Lehman, J. S., Santner, T. J., and Notz, W. I. (2004). Designing computer experiments to determine robust control variables. \textit{Statistica Sinica}, 14, 571--590.

\item{}
Marzat, J., Walter, E., and Piet-Lahanier, H. (2013). Worst-case global optimization of black-box functions through Kriging and relaxation. \textit{Journal of Global Optimization}, 55, 707--727.

\item{}
Marzat, J., Walter, E., and Piet-Lahanier, H. (2016). A new expected-improvement algorithm for continuous minimax optimization. \textit{Journal of Global Optimization}, In press.

\item{}
McKay, M. D., Beckman, R. J. and Conover, W. J. (1979). A comparison of three methods for selecting values of input variables in the analysis of output from a
computer code. \textit{Technometrics}, 21, 239--245.

\item{}
Niederreiter, H., (1992). \textit{Random Number Generation and Quasi-Monte Carlo Methods}, NSFCBMS Regional Conference on Random Number Generation. (Vol. 63). Philadelphia:
Society for Industrial and Applied mathematics.




\item{}
Picheny, V., Ginsbourger, D., Richet, Y., and Caplin, G. (2013), Quantile-based optimization of noisy computer experiments with tunable precision,
\textit{Technometrics}, 55, 2--13.




\item{}
Ruberg, S. J., Chen, L., and Wang, Y. (2010). The mean does not mean as much anymore: Finding sub-groups for tailored therapeutics. \textit{Clinical Trials}, 7, 574--583.

\item{}
Sacks, J., Welch, W. J., Mitchell, T. J. and Wynn, H. P. (1989). Design and analysis of computer experiments. \textit{Statistical Science}, {4}, 409--423.

\item{}
Saltelli, A., Ratto, M., Andres, T., Campolongo, F., Cariboni, J., Gatelli, D., Saisana, M. and Tarantola, S. (2008). \textit{Global Sensitivity Analysis: The Primer}. John Wiley \& Sons.

\item{}
Santner, T. J., Williams, B. J. and Notz, W. I. (2003). \textit{The Design and Analysis of Computer Experiments}. Springer, New York.


\item{}
Ugray, Z., Lasdon, L., Plummer, J., Glover, F., Kelly, J., and Marti, R. (2007). Scatter search and local NLP solvers: a multistart framework for global optimization. \textit{INFORMS Journal on Computing}, 19, 328--340.


\item{}
Williams, B. J., Santner, T. J., and Notz, W. I. (2000). Sequential design of computer experiments to minimize integrated response functions. \textit{Statistica Sinica}, 10, 1133--1152.


\item{}
Xu, Y., Yu, M., Zhao, Y. Q., Li, Q., Wang, S., and Shao, J. (2015). Regularized outcome weighted subgroup identification for differential treatment effects. \textit{Biometrics}, 71, 645--653.

\item{}
Zhao, Y. Q., Zeng, D., Laber, E. B., and Kosorok, M. R. (2015). New statistical learning methods for estimating optimal dynamic treatment regimes. \textit{Journal of the American Statistical Association}, 110, 583--598.

\end{description}}

\end{document}